\documentclass[journal]{IEEEtran}

\usepackage{amsmath,amssymb,latexsym, bm,bbm,mathrsfs}
\usepackage{epsfig}
\usepackage{caption}

\newlength{\noteWidth}
\long\def\notes#1{\ifinner
           {\tiny #1}
           \else
           \marginpar{\parbox[t]{\noteWidth}{\raggedright\tiny #1}}
       \fi\typeout{#1}}

       \newtheorem{thm}{\bf{Theorem}}[section]
       
       \newtheorem{lem}{\bf{Lemma}}[section]
       
       \newtheorem{prop}{\bf{Proposition}}[section]
       \newtheorem{defn}{\bf{Definition}}[section]
  \newtheorem{remark}{\bf{Remark}}[section]

\newcounter{mytempeqncnt}

\fussy

\begin{document}

\allowdisplaybreaks

\title{On Optimal Causal Coding of Partially Observed Markov Sources in Single and Multi-Terminal Settings}
\author{Serdar Y\"uksel$^1$}

       \maketitle
%

\footnotetext[1]{Department of Mathematics and Statistics, Queen's University, Kingston, Ontario, Canada, K7L 3N6. Research supported by the Natural Sciences and Engineering Research Council of Canada (NSERC). Email: yuksel@mast.queensu.ca. This work was presented in part at the at the IEEE International Symposium on Information Theory, Austin, Texas, in 2010.\\
Copyright (c) 2012 IEEE. Personal use of this material is permitted.  However, permission to use this material for any other purposes must be obtained from the IEEE by sending a request to pubs-permissions@ieee.org.}

\begin{abstract}
The optimal causal (zero-delay) coding of a partially observed Markov process is studied, where the cost to be minimized is a bounded, non-negative, additive, measurable single-letter function of the source and the receiver output. A structural result is obtained extending Witsenhausen's and Walrand-Varaiya's structural results on optimal causal coders to more general state spaces and to a partially observed setting. The decentralized (multi-terminal) setup is also considered.  For the case where the source is an i.i.d. process, it is shown that an optimal solution to the decentralized causal coding of correlated observations problem is memoryless. For Markov sources, a counterexample to a natural separation conjecture is presented.
\end{abstract}

\section{Introduction}

This paper considers optimal causal encoding/quantization of partially observed Markov processes. We begin with providing a description of the system model. We consider a partially observed Markov process, defined on a probability space $(\Omega,{\cal F},P)$ and described by the following discrete-time equations for $t \geq 0$:
\begin{eqnarray}
x_{t+1} &=& f(x_t, w_t), \label{sourceModel} \\
y^i_t&=&g^i(x_t,r^i_t), \label{channelModel}
\end{eqnarray}
for (Borel) measurable functions $f, g^i,i=1,2$, with $\{w_t, r^i_t, i=1,2\}$ i.i.d., mutually independent noise processes and $x_0$ a random variable with probability measure $\nu_0$. Here, we let $x_t \in \mathbb{X}$, and $y^i_t \in \mathbb{Y}^i$, where $\mathbb{X},\mathbb{Y}^i$ are complete, separable, metric spaces (Polish spaces), and thus, include countable spaces or $\mathbb{R}^n$, $n \in \mathbb{N}_+$.

\begin{figure}[h]
\centering
\epsfig{figure=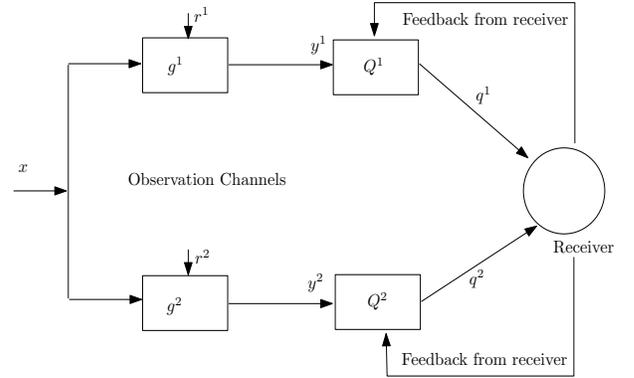,height=5cm,width=8cm}
\caption{Partially observed source under a decentralized structure. \label{LLL}}
\end{figure}

Let an encoder, Encoder $i$, be located at one end of an observation channel characterized by (\ref{channelModel}). The encoders transmit their information to a receiver (see Figure \ref{LLL}), over a discrete noiseless channel with finite capacity; that is, they quantize their information. The information at the encoders may also contain feedback from the receiver, which we clarify in the following.

Let us first define a quantizer.
\begin{defn}
Let ${\cal M} = \{1,2,\dots,M\}$ with $M=|{\cal M}|$. Let $\mathbb{A}$ be a topological space. A quantizer $Q(\mathbb{A};{\cal M})$ is a Borel measurable map from $\mathbb{A}$ to ${\cal M}$. \hfill $\diamond$
\end{defn}

When the spaces $\mathbb{A}$ and ${\cal M}$ are clear from context, we will drop the notation and denote the quantizer simply by $Q$.

We refer by a {\bf Composite Quantization Policy} $\Pi^{comp,i}$ of Encoder $i$, a sequence of functions $\{Q_t^{comp,i}, t \geq 0 \}$ which are causal such that the quantization output at time $t$, $q^i_t$, under $\Pi^{comp,i}$ is generated by a causally measurable function of its local information, that is, a mapping measurable on the sigma-algebra generated by
 \[I^i_t=\{y^i_{[0,t]}, q^1_{[0,t-1]}, q^2_{[0,t-1]} \}, \quad t \geq 1,\] and $I^i_0=\{y^i_0\},$ to a finite set ${\cal M}^i_t$, the quantization output alphabet at time $t$ given by \[{\cal M}^i_t := \{1,2,\dots, |{\cal M}^i_t|\},\]
for $0 \leq t \leq T-1$ and $i=1,2$. Here $\{{\cal M}^i_t\}$ are fixed in advance and do not depend on the realizations of the random variables. Here, we have the notation for $t \geq 1$:
\[y^i_{[0,t-1]} = \{y^i_s, 0 \leq s \leq t-1 \}.\]

Let \[\mathbb{I}^i_t = \bigg( \prod_{s=0}^{t-1}  {\cal M}^1_s \times {\cal M}^2_s  \bigg) \times (\mathbb{Y}^i)^{t+1}, \quad t \geq 1, \quad \quad \mathbb{I}^i_0 = \mathbb{Y}^i,\]
 be information spaces such that for all $t \geq 0$, $I^i_t \in \mathbb{I}^i_t$. Thus,
\[Q_t^{comp,i}: \mathbb{I}^i_t \to {\cal M}^i_t.\]

We may express, equivalently, the policy $\Pi^{comp,i}$ as a composition of a {\bf Quantization Policy} $\Pi^i$ and a {\bf Quantizer}. A quantization policy of encoder i, ${\cal T}^i$, is a sequence of functions $\{T^i_t\}$, such that for each $t \geq 0$, $T^i_t$ is a mapping from the information space $\mathbb{I}^i_t$ to a space of quantizers $\mathbb{Q}^i_t$. A quantizer, subsequently is used to generate the quantizer output. That is for every $t$ and $i$, $T^i_t(I_t) \in \mathbb{Q}^i_t$ and for every $I^i_t \in \mathbb{I}^i_t$, we will adopt a following representation
\begin{eqnarray}\label{duplicate}
Q_t^{comp,i}(I^i_t) = (T^i_t(I^i_t)) (I^i_t),
 \end{eqnarray}
 mapping the information space to ${\cal M}^i_t$ in its most general form. We note that even though there may seem to be duplicated information in (\ref{duplicate}) (since a map is used to pick a quantizer, and the quantizer maps the available information to outputs) we will eliminate any informational redundancy: A quantizer action will be generated based on the common information at the encoder and the receiver, and the quantizer will map the relevant private information at the encoder to the quantization output. Such a separation in the design will also allow us to use the machinery of Markov Decision Processes to obtain a structural method to design optimal quantizers, to be clarified further, without any loss in optimality.

That is, let the information at the receiver at time $t \geq 0$ be $I^r_t=\{q^1_{[0,t-1]},q^2_{[0,t-1]}\}$. The common information, under feedback information, in the encoders and the receiver is the set $I^r_t \in \bigg( \prod_{s=0}^{t-1}  {\cal M}^1_s \times {\cal M}^2_s  \bigg)$. Thus, we can express the composite quantization policy as:
\begin{eqnarray}\label{form2}
Q_t^{comp,i}(I^i_t) = (T^i_t(I^r_t)) (I^i_t \setminus I^r_t),
\end{eqnarray}
mapping the information space to ${\cal M}^i_t$. We note that, any composite quantization policy $Q_t^{comp,i}$ can be expressed in the form above; that is there is no loss in the space of possible such policies, since for any $Q_t^{comp,i}$, one could define \[T^i_t(I^r_t) (\cdot) := Q_t^{comp,i}(I^r_t, \cdot).\]
Thus, we let DM$^i$ have policy ${\cal T}^i$ and under this policy generate quantizer actions $\{ Q^i_t, t \geq 0 \}$, $Q^i_t \in \mathbb{Q}^i_t$ ($Q^i_t$ is the quantizer used at time $t$). Under action $Q^i_t$, and given the local information, the encoder generates $q^i_t$, as the {\em quantization output} at time $t$.

The receiver, upon receiving the information from the encoders, generates its decision at time $t$, also causally: An admissible causal receiver policy is a sequence of measurable functions $\gamma=\{\gamma_t\}$ such that $$\gamma_t : \prod_{s=0}^{t} \bigg( {\cal M}^1_s \times {\cal M}^2_s  \bigg) \to \mathbb{U}, \quad \quad t \geq 0$$
where $\mathbb{U}$ denotes the decision space.

For a general vector $a$, let ${\bf a}$ denote $\{a^1,a^2\}$ and let ${\bf \Pi} = \{\Pi^1, \Pi^2\}$ denote the ensemble of policies and ${\bf Q}_t=\{Q^1_t,Q^2_t\}$. Hence, ${\bf q}_{[0,t]}$ denotes $\{q^1_{[0,t]},q^2_{[0,t]}\}$. 

With the above formulation, the objective of the decision makers is the following minimization problem:
\begin{eqnarray*}\label{CostFunction}
\inf_{{\bf \Pi}^{comp}} \inf_{\gamma} E^{{\bf \Pi}^{comp},\gamma}_{\nu_0}[\sum_{t=0}^{T-1} c(x_t,v_t)],
\end{eqnarray*}
over all policies ${\bf \Pi}^{comp}$, $\gamma$ with the random initial condition $x_0$ having probability measure $\nu_0$. Here $c(\cdot,\cdot),$ is a non-negative, bounded, measurable function and $v_t=\gamma_t({\bf q}_{[0,t]})$ for $t \geq 0$.

We also assume that the encoders and the receiver know the apriori distribution $\nu_0$.

Before concluding this section, it may be worth emphasizing the operational nature of causality; as different approaches have been adopted in the literature. The encoders at any given time can only use their local information to generate the quantization outputs. The receiver, at any given time, can only use its local information to generate its decision/estimate. These happen with zero delay, that is if there is a common-clock at the encoders and the receiver; the receiver at time $t$ needs to make its decision before the realizations $x_{t+1},y^1_{t+1},y^2_{t+1}$ have taken place. This corresponds to the zero-delay coding schemes of, for example, Witsenhausen and Linder-Lugosi in \cite{Witsenhausen}, \cite{LinderLugosi}; but is different from the setup of Neuhoff and Gilbert \cite{NeuhoffGilbert}, which allows long delays at the decoder. Our motivation for such zero-delay schemes comes mainly from applications in networked control systems, when sensors need to transmit information to controllers who need to act on a system; such systems cannot tolerate long delays, in particular when the source is open loop unstable and disturbance exists in the evolution of the source.

\subsection{Relevant literature}

Some related studies of the above setup include optimal control with multiple sensors and sequential decentralized hypothesis testing problems and multi-access communications with feedback \cite{AchilleasAllerton09}. Related papers on real-time coding include the following: \cite{NeuhoffGilbert} established that the optimal optimal causal encoder minimizing the data rate subject to a distortion for an i.i.d sequence is memoryless. 
If the source is $k$th-order Markov, then the optimal causal fixed-rate coder minimizing any measurable distortion uses only the last $k$ source symbols, together with the current state at the receiver's memory \cite{Witsenhausen}. Reference \cite{WalrandVaraiya} considered the optimal causal coding problem of finite-state Markov sources over noisy channels with noiseless feedback. \cite{Teneketzis} and \cite{MahTen09} considered optimal causal coding of Markov sources over noisy channels without feedback. \cite{MahajanTeneketzisJSAC} considered the optimal causal coding over a noisy channel with noisy feedback. Reference \cite{LinderZamir} considered the causal coding of stationary sources under a high-rate assumption.

Our paper is particularly related to the following two efforts in the literature: Borkar, Mitter and Tatikonda \cite{Borkar} studies a related problem of coding of a partially observed Markov source, however, the construction for the encoders is restricted to take a particular form which uses the information at the decoder and the most recent observation at the encoder (not including the observation history). As another point of relevance with our paper, \cite{Borkar} regarded the actions as the {\em quantizer functions}, which we will discuss further. In contrast, the only restriction we have in this paper is causality, in the zero-delay sense. On the other hand, we do not claim the existence results that the authors in \cite{Borkar} are making. Another work in the literature which is related to ours is by Nayyar and Teneketzis \cite{NayyarTeneketzis}, considering a multi-terminal setup. \cite{NayyarTeneketzis} considers decentralized coding of correlated sources when the encoders observe conditionally independent messages given a finitely valued random variable and obtain separation results for the optimal encoders. The paper also considers noisy channels. In our setup, there does not exist a finitely valued random variable which makes the observations at the encoders conditionally independent.


References \cite{Weissman} and \cite{KaspiMerhav} consider optimal causal variable-rate coding under side information and \cite{YukBasMeynITA08} considers optimal variable-rate causal coding under distortion constraints. The studies in \cite{KaspiMerhav} and \cite{YukBasMeynITA08} are in the context of real-time, zero-delay settings; whereas \cite{Weissman} considers causality in the sense of Neuhoff and Gilbert \cite{NeuhoffGilbert} as discussed in the previous section.

We will also obtain structural results for optimal decentralized coding of i.i.d. sources. There are algorithmic results available in the literature when the encoders satisfy the optimal structure obtained in the paper, important resources in this direction include \cite{Gray1}, \cite{Gray2} and \cite{Skoglund}.

A parallel line of consideration which has a rate-distortion theoretic nature is on {\em sequential-rate distortion} proposed in \cite{Tatikonda} and the {\em feedforward} setup, which has been investigated in \cite{Feedforward2} and \cite{Feedforward1}.

Our work is also related to Witsenhausen's {indirect rate distortion} problem \cite{WitsenhausenIndirect} (see also \cite{Tsybakov}). We will observe that, the separation argument through the {\em disconnection principle} of \cite{WitsenhausenIndirect} applies to our setting in a dynamic context. Further related papers include \cite{BansalBasarSysCont}, \cite{HuangDey}.

In our paper, we also use ideas from team decision theory, see \cite{WitsenhausenIntrinsic}, \cite{YukTAC09}, \cite{Mahajan}, \cite{MahajanCDC2011} and \cite{ComoYukConCom} for related discussions and applications.


\subsection{Contributions of the paper}

\begin{itemize}
\item The optimal causal coding of a partially observed Markov source is considered. For the single terminal case, a structural result is obtained extending Witsenhausen's and Walrand and Varaiya's structural results on optimal causal (zero-delay) coders to a partially observed setting and to sources which take values in a Polish space. We show that a separation result of a form involving {the decoder's belief on the encoder's belief on the state} is optimal.

\item The decentralized (multi-terminal) setup is also considered. For the case where the source is an i.i.d. process, it is shown that the optimal decentralized causal coding of correlated observations problem admits a solution which is memoryless. For Markov sources, a counterexample to a natural separation conjecture is presented. The decentralized control concept of {\em signaling} is interpreted in the context of decentralized coding.

    \item The results are applied to a Linear-Quadratic-Gaussian (LQG) estimation/optimization problem. The results above induce an optimality result for {\em separation of estimation and quantization}, where the estimation is obtained with a Kalman Filter and the filter output is quantized.
    \end{itemize}

We now summarize the rest of the paper. In section II, we present our results on optimal coding of a partially observed Markov process when there is only one encoder. Section III discusses the decentralized setting for a multi-encoder setup and presents a counterexample for a separation conjecture and provides a separation result when the source is memoryless. The paper ends with the concluding remarks of section V, following an application example on linear, Gaussian systems in Section IV. The proofs of the results are presented in the Appendix.

\section{Single-Terminal Case: Optimal Causal Coding of a Partially Observed Markov Source}

\subsection{Revisiting the single-terminal, fully observed case}

Let us revisit the single-encoder, fully observed case: In this setup, $y_t=x_t$ for all $t \geq 0$. There are two related approaches in the literature as presented explicitly by Teneketzis in \cite{Teneketzis}; one adopted by Witsenhausen \cite{Witsenhausen} and one by Walrand and Varaiya \cite{WalrandVaraiya}. Reference \cite{Teneketzis} extended the setups to the more general context of non-feedback communication.

\begin{thm}\label{WitsenTheoremFull}[Witsenhausen \cite{Witsenhausen}]
Any {\em (causal) composite quantization policy} can be replaced, without any loss in performance, by one which only uses $x_t$ and $q_{[0,t-1]}$ at time $t \geq 1$. \hfill $\diamond$
\end{thm}

Walrand and Varaiya considered sources living in a finite set, and obtained the following:

\begin{thm}\label{WalVarTheoremFull}[Walrand-Varaiya \cite{WalrandVaraiya}]
Any optimal {\em (causal) composite quantization policy} can be replaced, without any loss in performance, by one which only uses the conditional probability measure $P(x_{t}|q_{[0,t-1]})$, the state $x_t$, and the time information $t$, at time $t \geq 1$. \hfill $\diamond$ 
\end{thm}

The difference between the structural results above is the following: In the setup suggested by Theorem \ref{WitsenTheoremFull}, the encoder's memory space is not fixed and keeps expanding as the decision horizon in the optimization, $T-1$, increases. In Theorem \ref{WalVarTheoremFull}, the memory space of an optimal encoder is fixed. In general, the space of probability measures is a very large one; however, it may be the case that different quantization outputs may lead to the same conditional probability measure on the state process, leading to a reduction in the required memory. Furthermore, Theorem \ref{WalVarTheoremFull} allows one to apply the theory of Markov Decision Processes, for infinite horizon problems. We note that \cite{Borkar} applied such a machinery to obtain existence results for optimal causal coding of partially observed Markov processes


\subsection{Optimal causal coding of a partially observed Markov source}


Consider the setup earlier in (\ref{channelModel}) with a single encoder. Thus, the system considered is a discrete-time scalar system described by
\begin{eqnarray}\label{ProblemModel4}
x_{t+1}=f(x_{t},w_t), \quad \quad y_t= g(x_t,r_t),
\end{eqnarray}
 where $x_t$ is the state at time $t$, and $\{w_t, r_t \}$ is a sequence of zero-mean, mutually independent, identically distributed (i.i.d.) random variables with finite second moments. 
Let the quantizer, as described earlier, map its information to a finite set ${\cal M}_t$.
At any given time, the receiver generates a quantity $v_t$ as a function of its received information, that is as a function of $\{q_0,q_1,\dots,q_t\}$. The goal is to minimize $\sum_{t=0}^{T-1} E[c(x_t,v_t)],$ subject to constraints on the number of quantizer bins in ${\cal M}_t$, and the causality restriction in encoding and decoding.



Let for a Polish space $\mathbb{S}$, ${\cal P}(\mathbb{S})$ be the space of probability measures on ${\cal B}(\mathbb{S})$, the Borel $\sigma-$field on $\mathbb{S}$ (generated by open sets in $\mathbb{S}$). At this point pause to provide a brief discussion on the space ${\cal P}(\mathbb{S})$.

Let $\Gamma(\mathbb{S})$ be the set of all Borel measurable and bounded functions from $\mathbb{S}$ to $\mathbb{R}$. We first wish to find a topology on ${\cal P}(\mathbb{S})$, under which functions of the form:
\[ \Theta:= \{ \int_{x\in \mathbb{S}} \pi(dx) f(x), \quad f \in \Gamma(\mathbb{X}) \} \]
 are measurable on ${\cal P}(\mathbb{S})$. We will need this to construct the structure of optimal quantizers later in this section.

 Let $\{\mu_n,\, n\in \mathbb{N}\}$ be a sequence in
$\mathcal{P}(\mathbb{S})$. Recall that
$\{\mu_n\}$ is said to  converge
  to $\mu\in \mathcal{P}(\mathbb{S})$ \emph{weakly} if
\[
 \int_{\mathbb{S}} c(x) \mu_n(dx)  \to \int_{\mathbb{S}}c(x) \mu(dx)
\]
for every continuous and bounded $c: \mathbb{S} \to \mathbb{R}$.
The sequence $\{\mu_n\}$ is said to  converge
  to $\mu\in \mathcal{P}(\mathbb{S})$  \emph{setwise} if
\[
 \int_{\mathbb{S}} c(x) \mu_n(dx)  \to \int_{\mathbb{S}} c(x) \mu(dx)
\]
for every measurable and bounded $c: \mathbb{S} \to
\mathbb{R}$.
For two probability measures $\mu,\nu \in
\mathcal{P}(\mathbb{S})$, the \emph{total variation} metric
is given by
\begin{eqnarray}
\|\mu-\nu\|_{TV}&:= & 2 \sup_{B \in {\cal B}(\mathbb{S})}
|\mu(B)-\nu(B)| \nonumber \\
 &=&  \sup_{f: \, \|f\|_{\infty} \leq 1} \bigg| \int f(x)\mu(dx) -
\int f(x)\nu(dx) \bigg|, \nonumber
\end{eqnarray}
where the infimum is over all measurable real $f$ such that
$\|f\|_{\infty} = \sup_{x \in \mathbb{S}} |f(x)|\le 1$.
A sequence  $\{\mu_n\}$ is said to  converge
  to $\mu\in \mathcal{P}(\mathbb{S})$ in total variation if
$\| \mu_n - \mu   \|_{TV}  \to 0.$

These  three convergence notions are in increasing order of strength: convergence in
total variation implies setwise convergence, which in turn implies
weak convergence. Total variation is a very strong notion for convergence. Furthermore, the space of probability measures under total variation metric is not separable. Setwise convergence also induces an inconvenient topology on the space of probability measures, particularly because this topology is not metrizable (\cite[p.~59]{Ghosh}). However, the space of probability measures on a complete, separable, metric (Polish) space endowed with the topology
of weak convergence is itself a complete, separable, metric space
\cite{InfiniteDimensionalAnalysis}. The Prohorov metric, for example, can be used to
metrize this space, among other metrics. This topology has found many applications in
information theory and stochastic control. For these reasons, one would like to work with weak convergence.

  By the above definitions, it is evident that both setwise convergence and total variation are sufficient for measurability of the function class $\Theta$, since under these topologies $\int \pi(dx) f(x)$ is (sequentially) continuous on ${\cal P}(\mathbb{S})$ for every $f \in \Gamma(\mathbb{S})$. However, as we state in the following, weak convergence is also sufficient (see Theorem 15.13 in \cite{InfiniteDimensionalAnalysis} or p. 215 in \cite{Bogachev}).
\begin{thm}\label{MonotoneClassRamon}
Let $\mathbb{S}$ be a Polish space and let $M(\mathbb{S})$ be the set of all measurable and bounded functions $f: \mathbb{S} \to \mathbb{R}$ under which \[\int \pi(dx) f(x)\] defines a measurable function on ${\cal P}(\mathbb{S})$ under the weak convergence topology.
Then, $M(\mathbb{S})$ is the same as $\Gamma(\mathbb{S})$ of all bounded and measurable functions. \hfill $\diamond$
\end{thm}

Hence, ${\cal P}(\mathbb{S})$ will denote the space of probability measures on $\mathbb{S}$ under weak convergence. Now, define $\pi_t \in {\cal P}(\mathbb{X})$ to be the regular conditional probability measure given by
 $$\pi_t(A) = P(x_t \in A|y_{[0,t]}), \quad A \in {\cal B}(\mathbb{X}).$$
The existence of this regular conditional probability measure for every realization $y_{[0,t]}$ follows from the fact that both the state process and the observation process are Polish. It is known that the process $\{\pi_t\}$ evolves according to a non-linear filtering equation (see (\ref{MarkovBelief})), and is itself a Markov process (see \cite{BorkarBudhiraja}, \cite{Stettner}, \cite{Borkar}).

Let us also define $\Xi_t \in {\cal P}({\cal P}(\mathbb{X}))$ as the regular conditional measure
$$\Xi_t(A) = P(\pi_t \in A |q_{[0,t-1]}), \quad A \in {\cal B}({\cal P}(\mathbb{X})).$$

The following are the main results of this section:

\begin{thm}\label{thm1}
Any {\em (causal) composite quantization policy} can be replaced, without any loss in performance, by one which only uses $\{\pi_t, q_{[0,t-1]} \}$ as a sufficient statistic for $t \geq 1$. This can be expressed as a quantization policy which only uses $q_{[0,t-1]}$ to generate a quantizer, where the quantizer uses $\pi_t$ to generate the quantization output at time $t$. \hfill $\diamond$
\end{thm}
%

\begin{thm}\label{thm3}
Any optimal {\em (causal) composite quantization policy} can be replaced, without any loss in performance, by one which only uses $\{\Xi_t,\pi_t,t\}$ for $t \geq 1$. This can be expressed as an optimal quantization policy which only uses $\{\Xi_t,t\}$ to generate an optimal quantizer, where the quantizer uses $\pi_t$ to generate the quantization output at time $t$. \hfill $\diamond$
\end{thm}

The proofs of the results above are presented in the Appendix. We present two remarks in the following:

\begin{remark}
Our results above are not surprising. In fact, once one recognizes the fact that $\{\pi_t\}$ forms a Markov source, and the cost function can be expressed as a function $\tilde{c}(\pi,v)$, for some function $\tilde{c}: {\cal P}(\mathbb{X}) \times \mathbb{U} \to \mathbb{R}$, one could almost directly apply Witsenhausen's \cite{Witsenhausen} as well as Walrand and Varaiya's \cite{WalrandVaraiya} results to recover the structural results above (except the fact that Walrand and Varaiya consider sources living in a finite alphabet). The proofs in the Appendix are presented for completeness and to address the technical intricacies. \hfill $\diamond$ \end{remark}
\begin{remark}
Having the actions as the quantizers, and not the quantizer outputs, allows one to define a Markov Decision Problem with well-defined cost functions and state and action spaces. By the proof of Theorem \ref{thm3}, we will observe that $(\Xi_,Q_t)$ forms a Markov Chain, which is a key observation: Thus, the action space can be constructed as some topological space of quantizers acting on ${\cal P}({\mathbb{X}})$. Borkar, Mitter and Tatikonda \cite{Borkar} adopted this view while formulating an MDP optimization problem, where the quantizer acts on $\mathbb{Y}$ (As mentioned earlier, our separation result is different from \cite{Borkar} due to the structure imposed on the quantizers in \cite{Borkar}.). See also \cite{YukselOptimizationofChannels} and \cite{YukLinCDC2012} for a topology on quantizers. \hfill $\diamond$
\end{remark}

\subsection{Extensions to finite delay decoding and higher order Markov sources}

The results presented are also generalizable to settings where, a) the source is Markov of order $m>0$, b) a finite delay $d$ is allowed at the decoder, and c) the observation process depends also on past source outputs in a sense described in (\ref{ProblemModel114}) below. For these cases, we consider the following generalization of the source by expanding the state space.

Suppose that the partially observed source is such that, the source is Markov of order $m$, or there is a finite delay $d>0$ which is allowed at the decoder. In this case, we can augment the source to obtain $z_t = \{x_{[t-\max(d+1,m)+1,t]}\}$. Note that $\{z_t\}$ is Markov. We can thus consider the following representation:
\begin{eqnarray}\label{ProblemModel114}
z_{t+1}=f(z_{t},w_t), \quad \quad y_t= g(z_t,r_t),
\end{eqnarray}
where $z_t = \{x_{[t-\max(d+1,m)+1,t]}\} \in \mathbb{X}^{\max(d+1,m)}$, and $r_t, w_t$ are mutually independent, i.i.d. processes.

Any per-stage cost function of the form $c(x_t,v_t)$ can be written as for some $\tilde{c}$: $\tilde{c}(z_t,v_t)$. For the finite delay case, the cost per-stage can further be specialized as $\tilde{c}(x_{t-d}, v_t)$. For the Markov case with memory, the cost function per-stage writes as $\tilde{c}(x_{[t-m+1,t]}, v_t)$. Now, by replacing $\mathbb{X}$ with $\mathbb{X}^{\max(d+1,m)}$, let $\pi_t \in {\cal P}(\mathbb{X}^{\max(d+1,m)})$ be given by
\[\pi_t(A) = P(z_t \in A | y_{[0,t]}), \quad A \in {\cal B}(\mathbb{X}^{\max(d+1,m)})\]
 and $\Xi_t \in {\cal P}({\cal P}(\mathbb{X}^{\max(d+1,m)}))$ be the regular conditional measure defined by \[\Xi_t(A) = P(\pi_t \in A |q_{[0,t-1]}), \quad A \in {\cal B}({\cal P}(\mathbb{X}^{\max(d+1,m)})).\]

Hence, we have the following result, which is a direct extension of Theorems \ref{thm1} and \ref{thm3}.

\begin{thm}\label{thm15}
Suppose that the partially observed source is such that, the source is Markov of order $m$, or there is a finite delay $d>0$ which is allowed at the decoder. With $z_t = \{x_{[t-\max(d+1,m)+1,t]} \}$, $y_t$ satisfies (\ref{ProblemModel114}). Then, we have the following extensions: \\
i) Any {\em causal composite quantization policy} can be replaced, without any loss in performance, by one which only uses $\{\pi_t, q_{[0,t-1]} \}$ as a sufficient statistic for $t \geq 1$. This can be expressed as a quantization policy which only uses $q_{[0,t-1]}$ to generate a quantizer, where the quantizer uses $\pi_t$ to generate the quantization output at time $t$. \\
ii) Any optimal {\em causal composite quantization policy} can be replaced, without any loss in performance, by one which only uses $\{\Xi_t,\pi_t,t\}$ for $t \geq 1$. This can be expressed as an optimal quantization policy which only uses $\{\Xi_t,t\}$ to generate an optimal quantizer, where the quantizer uses $\pi_t$ to generate the quantization output at time $t$. \hfill $\diamond$
\end{thm}

For a further case where the decoder's memory is limited or imperfect, the results apply by replacing the full information at the receiver considered so far in our analysis with the limited memory under additional technical assumptions on the decoder's update of its memory (in particular, (\ref{whySeparationHolds}) in the proof of Theorem \ref{thm3} does not apply in general). However, an equivalent result of Theorem \ref{thm1} applies also for the limited memory setting. Such memory settings have been considered in \cite{Witsenhausen}, \cite{WalrandVaraiya} and \cite{MahajanTeneketzisJSAC}.

\section{Multi-Terminal (Decentralized) Setup}

\subsection{Case with memoryless sources}

Let us first consider a special, but important, case when $\{x_t, t \geq 0\}$ is an independent and identically distributed (i.i.d.) sequence. Further, suppose that, the observations are given by
\begin{eqnarray}
y^i_t&=&g^i(x_t,r^i_t), \label{channelModel3}
\end{eqnarray}
for measurable functions $g^i,i=1,2$, with $\{r^1_t, r^2_t\}$ (across time) an i.i.d. noise process. We do not require that $r^1_t$ and $r^2_t$ are independent for a given $t$. We note that our result below is also applicable when the process $\{r^1_t, r^2_t\}$ is only independent (across time), but not necessarily identically distributed.

One difference with the general setup considered earlier in Section I is that we require the observation spaces $\mathbb{Y}^i, i=1,2$, to be finite spaces ($\mathbb{X}$ can still be Polish).

Suppose the goal is again the minimization problem
\begin{eqnarray}\label{problemFormulation2}
\inf_{{\bf \Pi}^{comp}} \inf_{\gamma} E^{{\bf \Pi}^{comp},\gamma}_{\nu_0}[\sum_{t=0}^{T-1} c(x_t,v_t)],
\end{eqnarray}
over all causal coding and receiver decision policies.

We now make a definition. In the following $1_E$ denotes the indicator function of an event $E$.
\begin{defn}
We define the class of non-stationary memoryless team policies at $t \geq 0$ as follows:
\begin{eqnarray}\label{piNSM}
&&\Pi^{NSM}:= \bigg\{{\bf \Pi}^{comp} : P({\bf q}_t | {\bf y}_{[0,t]}) \nonumber \\
&& = P(q^1_t | y^1_t,t)P(q^2_t | y^2_t,t) = 1_{\{q^1_t=Q^1_t(y^1_t)\}}1_{\{q^2_t=Q^2_t(y^2_t)\}}, \nonumber \\
&& \quad \quad \quad \quad \quad \quad  \quad Q^i_t: \mathbb{Y}^i \to {\cal M}^i_t, i=1,2, t \geq 0 \bigg\},
\end{eqnarray}
where, in the above, $\{Q^1_t,Q^2_t\}$ are arbitrary measurable functions.
\end{defn}

\begin{thm}\label{mainDecentralized}
Consider the minimization of (\ref{problemFormulation2}).
An optimal composite quantization policy over all causal policies is an element of $\Pi^{NSM}$. Such a policy exists. \hfill $\diamond$
\end{thm}

The proof is presented in the Appendix.

Hence, an optimal composite quantization policy only uses the product form admitted by a non-stationary memoryless team policy. It ignores the past observations and past quantization outputs. We note that, the proof also applies to the case when the source is memoryless, but not necessarily i.i.d..
One may ask why feedback could be useful when the source is i.i.d.. Feedback may be useful for at least two reasons: (i) Feedback can be used as a signaling mechanism for the encoders to communicate with each other (which we discuss further in the next subsections), and (ii) Feedback can provide common randomness to allow a convexification of the space of possibly randomized decentralized encoding strategies. Consider the optimization problem discussed in (\ref{problemFormulation2})
\[ J({\bf \Pi}^{comp}) =  \inf_{\gamma} E^{{\bf \Pi}^{comp},\gamma}_{\nu_0}[\sum_{t=0}^{T-1} c(x_t,v_t)]. \]
The function $J({\bf \Pi})$ is concave in the choice of a team policy ${\bf \Pi}$ (see Theorem 4.1 in \cite{YukLinAllerton2010} for the case with $T=1$, the proof also holds for the current setting). As such, if the space of joint encoding policies is convexified by common randomness, an optimal solution would exist at an extreme point; which in turn does not require a use of common randomness. This explains why common randomness generated by past quantization outputs does not present further benefit in the current setting.
%
%
%
%
%
%
%
%

We note before ending this subsection that if there is an entropy constraint on the quantizer outputs, then feedback might be useful for finite horizon problems as it provides common randomness, which cannot be achieved by time-sharing in a finite-horizon problem. \cite{NeuhoffGilbert} observed that randomization of two scalar quantizers (operationally achievable through time-sharing) is optimal in causal coding of an i.i.d. source subject to distortion constraints, which also applies in the side information setting of \cite{Weissman}. On the other hand, for the zero-delay setting, when one considers the distortion minimization problem subject to an entropy constraint, \cite{GyorgyLinder} observed that the distortion-entropy curve is non-convex, leading to a benefit of common-randomness for achieving points in the lower convex hull of this curve. Further relevant discussions on randomization and optimal quantizer design are present in \cite{GaborGyorfi} and \cite{YukselOptimizationofChannels}.

\subsection{Case with Markov sources: A counterexample with signaling}

We now consider Markov sources and exhibit that it is, in general, not possible to obtain a separation result of the form presented for the single-terminal case.

We will consider a two-encoder setup for the following result, where the encoders have access to the feedback from the receiver (Fig. \ref{LLL}). We have the following result.

\begin{prop}\label{NegativeResult}
Consider the setup in (\ref{sourceModel})-(\ref{channelModel}) and let $\pi^i_t=P(x_t|y^i_{[0,t]}), x_t \in \mathbb{X}$, $i=1,2$. 
An optimal composite quantization policy cannot, in general, be replaced by a policy which only uses $\{{\bf q}_{[0,t-1]},\pi^i_t\}$ to generate $q^i_t$ for $i=1,2$. \hfill $\diamond$
\end{prop}

\textbf{Proof.} It suffices to produce an instance where an optimal policy cannot admit the separated structure. Toward this end, let $z_1,z_2,z_3$ be uniformly distributed, independent, binary numbers; $x_0,x_1$ be defined by:
$$x_0= \begin{bmatrix} z_1 \\ z_2 \\ 0  \\ 0 \end{bmatrix}, \quad x_1= \begin{bmatrix} 0 \\ 0 \\ z_2 \\ z_3 \end{bmatrix},$$
such that $x_0(1)=z_1, x_0(2)=z_2, x_0(3)=x_0(4)=0$. Let the observations be given as follows:
\[y^1_t= g^1(x_t) = x_t(1) \oplus x_t(3) \oplus x_t(4) \]
\[ y^2_t= g^2(x_t) = x_t(1) \oplus x_t(2), \quad t=0,1.\]
where $\oplus$ is the x-or operation. That is,
\[y^1_0=\begin{bmatrix} z_1 \end{bmatrix}, \quad y^2_0=\begin{bmatrix} z_1 \oplus z_2\end{bmatrix}, \]
\[y^1_1= \begin{bmatrix}  z_2 \oplus z_3 \end{bmatrix}, \quad y^2_1= \begin{bmatrix} 0  \end{bmatrix} \]
Let the cost be:
\[E\bigg[(x_0(4) - E[x_0(4) | {\bf q}_{[0]}])^2 + (x_1(4) - E[x_1(4) | {\bf q}_{[0,1]}])^2\bigg]\]
That is, the cost is $E[(z_3 - E[z_3 | {\bf q}_{[0,1]}])^2]$, where $q^i_t$ are the information bits sent to the decoder for $t=0$ and $1$.

We further restrict the information rates to satisfy: $|{\cal M}^1_0|=|{\cal M}^1_1|=|{\cal M}^2_1|=2$, $|{\cal M}^2_0|=1$. That is, the encoder 2 may only send information at time $t=1$.

Under arbitrary causal composite quantization policies, a cost of zero can be achieved as follows: If the encoder 1 sends the value $z_1 $ to the receiver, and at time $1$, encoder 1 transmits $z_2 \oplus z_3$ and encoder 2 transmits $z_2$ (or $z_1 \oplus z_2$), the receiver can uniquely identify the value of $z_3$, for every realization of the random variables.

For such a source, an optimal composite policy cannot be written in the separated form, that is, an optimal policy of encoder 2 at time $1$ cannot be written as $h_1({\bf q}_{0},\pi^2_1)$, for some measurable function $h_1$. To see this, note the following: The conditional distribution on $x_1$ at encoder 2 at time $1$ is such that the conditional measure on $(z_2,z_3)$ is uniform and independent, that is $P(z_2=a,z_3=b|z_1 \oplus z_2)=(1/4)$ for all values of $a,b$. If a policy of the structure of $h_1$ is adopted, then it is not possible for encoder 2 to recall its past observation to extract the value of $z_2$. This is because, $\pi^2_1$ will be a distribution only on $z_2$ and $z_3$, which will be uniform and independent, given $z_1 \oplus z_2$. Thus, the information $y^2_0$ will not be available in the memory and the receiver will have access to at most $z_2 \oplus z_3$ and $z_1$ and $P(z_2, z_3 | z_1 \oplus z_2)$ (the last variable containing no useful information). The optimal estimator will be $E[z_3]=1/2$, leading to a cost of $1/4$.

\hfill $\diamond$

\subsection{Discussion: Connections with team decision theory}

In this subsection, we interpret the results of the previous subsections. We first provide a brief discussion on information structures in a decentralized optimization problem: Consider a collection of decision makers (DMs) where each has access to some local information variable. Such a collection of decision makers who wish to minimize a common cost function and who has an agreement on the system (that is, the probability space on which the system is defined, and the policy and action spaces) is said to be a {\em team}. Such a team is {\em dynamic} if the information of one DM is affected by the policy of some other DM. If there is a pre-specified order of action for the DMs, the team is said to be {\em sequential}. Witsenhausen \cite{WitsenhausenIntrinsic} provided the following characterization of information structures in a dynamic sequential team: Under a {\em centralized information structure}, all DMs have the same information. If a DM's, say DM$^j$, information is dependent on the policy of another DM, say DM$^k$, and DM$^j$ does not have access to the information available to DM$^k$, this information structure is said to admit a {\em non-classical information structure}. A decentralized system admits a {\em quasi-classical information structure}, if it is not non-classical.

In a decentralized optimization problem, when the information structure is non-classical, DMs might choose to communicate via their control actions: This is known as {\em signaling} in decentralized control (see for example \cite{YukTAC09}).

With the characterization of information structures above, every lossy coding problem is non-classical, since a receiver cannot recover the information available at the encoder fully, while its information is clearly affected by the coding policy of the encoder. However, in an encoding problem, the problem itself is the transmission of information. Therefore, we suggest the following: {\em Signaling in a coding problem is the policy of an encoder to use the quantizers/encoding functions to transmit a message to other decision makers, or to itself to be used in future stages, through the information sent to the receiver.} In the information theory literature, signaling has been employed in coding for Multiple Access Channels with feedback in \cite{CoverLeung}, \cite{BrossLapidoth} and \cite{Sandeep}. In these papers, the authors used active information transmission to allow for coordination between encoders.

The reason for the negative conclusion in Proposition \ref{NegativeResult} is that in general for an optimal policy,
\begin{eqnarray}\label{filtre}
P(q^i_{t} | \pi^i_{t}, {\bf q}_{[0,t-1]}, y^i_{[0,t-1]})  \neq P(q^i_{t} | \pi^i_{t}, {\bf q}_{[0,t-1]}),
\end{eqnarray}
when the encoders have engaged in signaling (in contrast with what we will have in the proof of the separation results). The encoders may benefit from using the received past observation variables explicitly.

Separation results for such dynamic team problems typically require information sharing between the encoders (decision makers), where the shared information is used to establish a sufficient statistic living in a fixed state space and which admits a controlled Markov recursion (hence, such a sufficient statistic can serve as a {\em state} for the decentralized system). For the proof of Theorem \ref{thm3}, we see that $\Xi_t$ forms such a state. For the proof of Theorem \ref{mainDecentralized}, we see that information sharing is not needed for the encoders to agree on a sufficient statistic, since the source considered is memoryless. Furthermore, for the multi-terminal setting with a Markov source, a careful analysis of the proof of Theorem \ref{mainDecentralized} (see (\ref{longequationDecentralized}), (\ref{BeliefOnBelief}) and the subsequent discussion) reveals that if the encoders agree on $P(dx_t|{\bf y}_{[0,t-1]})$ through sharing their beliefs for $t \geq 1$, then a separation result involving this joint belief can be obtained. See \cite{YukTAC09} for a related information sharing pattern and discussions. Further results on such a {\em dynamic programming approach to dynamic team problems} are present in \cite{AicardiDavoli}, \cite{Kurtaran}, \cite{NayyarMahajanTeneketzis}, \cite{MahajanCDC2011}, among other references.

\begin{remark} In the context of multi-terminal systems, for the computation of the capacity of Multiple Access Channels with memory and partial state feedback at the encoders, a relevant discussion has been reported in \cite{ComoYukConCom} (section V). This is in the same spirit as our current paper in that, \cite{ComoYukConCom} obtains an optimality result when the channel is memoryless; and points out the difficulties arising in the case of channels with memory in view of intractability of the optimization problem: One cannot identify a finite dimensional sufficient statistic for the encoders to use. Along a relevant direction, Section III.D of \cite{NayyarTeneketzis}, in the context of real-time coding, discusses the issue of the growing state space. \hfill $\diamond$
\end{remark}

\section{Application to Linear Quadratic Gaussian (LQG) Estimation Problems}

\begin{figure}[h]
\centering
\epsfig{figure=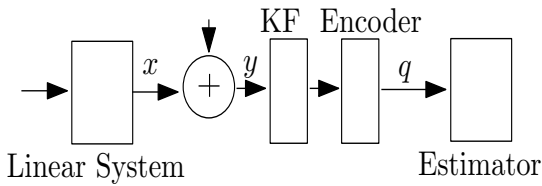,height=2.8cm,width=8cm}
\caption{{\bf Separation of Estimation and Quantization}: When the source is Gaussian, generated by the linear system (\ref{channelModel2}), the cost is quadratic, and the observation channel is Gaussian, the separated structure of the encoder above is optimal. That is, first the encoder runs a Kalman Filter (KF), and then causally encodes its estimate. For one-shot and independent observations setups, this result was observed in \cite{Ayanoglu}, \cite{Berger}, \cite{BansalBasarSysCont}, \cite{Tsybakov}, and \cite{Fine}. Our result shows that, an extension of this result applies for the optimal causal encoding of partially observed Markov sources as well. \label{KalmanOptimal}}
\end{figure}

Consider a Linear Quadratic Gaussian setup, where a sensor quantizes its noisy information to an estimator. Let $x_t \in \mathbb{R}^n, y_t \in \mathbb{R}^m$, and the evolution of the system be given by the following:
\begin{eqnarray}
x_{t+1} &=& Ax_t + w_t, \nonumber \\
y_t&=& Cx_t + r_t, \label{channelModel2}
\end{eqnarray}
Here, $\{w_t,r_t\}$ is a mutually independent, zero-mean Gaussian noise sequence with $W=E[w_tw_t'], R=E[r_tr_t']$ (where for a vector $w$, $w'$ denotes its transpose), $x_0$ is a zero mean Gaussian variable, and $A,C$ are matrices of appropriate dimensions. Suppose the goal is the computation of
\begin{eqnarray}\label{LQGopt}
\inf_{\Pi^{comp}} \inf_{\gamma} E^{\Pi^{comp},\gamma}_{\nu_0}[\sum_{t=0}^{T-1} (x_t-v_t)'Q(x_t-v_t)],
\end{eqnarray}
with $\nu_0$ denoting a Gaussian distribution for the initial state, $Q > 0$ a positive definite matrix (See Figure \ref{KalmanOptimal}).

The conditional measure $\pi_t = P(dx_t|y_{[0,t]})$ is Gaussian for all time stages, which is characterized uniquely by its mean and covariance matrix for all time stages. We have the following.

 \begin{thm}\label{LQGApplication}
For the minimization of the cost in (\ref{LQGopt}), any {\em causal composite quantization policy} can be replaced, without any loss in performance, by one which only uses the output of the Kalman Filter and the information available at the receiver. \hfill $\diamond$
\end{thm}

\textbf{Proof.}
The result can be proven by considering a direct approach, rather than as an application of Theorems \ref{thm1} and \ref{thm3} (which require bounded costs, however, this assumption can be relaxed for this case), exploiting the specific quadratic nature of the problem. Let $||\cdot||_Q$ denote the norm generated by an inner product of the form: $\langle x, y \rangle = x'Qy$ for $x,y \in \mathbb{R}^n$ for positive-definite $Q > 0$. The Projection Theorem for Hilbert Spaces implies that the random variable $x_t - E[x_t|y_{[0,t]}]$ is orthogonal to the random variables $\{y_{[0,t]},q_{[0,t]}\}$, where $q_{[0,t]}$ is included due to the Markov chain condition \[P(dx_t|y_{[0,t]},q_{[0,t]})=P(dx_t|y_{[0,t]}).\] We thus obtain the following identity.
\begin{eqnarray}\label{hilbert}
&& E[\|x_t - E[x_t|q_{[0,t]}]\|_Q^2]  = E[\|x_t - E[x_t|y_{[0,t]}]\|_Q^2]  \nonumber \\
&&  \quad  \quad + E\bigg[\bigg\|E[x_t|y_{[0,t]}] - E\bigg[E[x_t|y_{[0,t]}] \bigg| q_{[0,t]} \bigg] \bigg\|_Q^2\bigg]. \nonumber
\end{eqnarray}
The second term is to be minimized through the choice of the quantizers. Hence, the term $\tilde{m}_t:=E[x_t|y_{[0,t]}]$, which is computed through a Kalman Filter, is to be quantized (see Figure \ref{KalmanOptimal}). Recall that by the Kalman Filter (see \cite{Kushner}) with
\[\Sigma_{0|-1}=E[x_0x_0']\]
and for $t \geq 0$,
\begin{eqnarray}
&& \Sigma_{t+1|t} = A \Sigma_{t|t-1} A' + W \nonumber \\
&& \quad \quad  - (A \Sigma_{t|t-1}C') (C \Sigma_{t|t-1} C' + R)^{-1} (C \Sigma_{t|t-1} A'), \nonumber
\end{eqnarray}
the following recursion holds for $t \geq 0$ and with $\tilde{m}_{-1}=0$:
\begin{eqnarray} \label{POMDPtoFOMDPGaussian}
&& \tilde{m}_t=A\tilde{m}_{t-1} \nonumber \\
&&  + \Sigma_{t|t-1}C' (C \Sigma_{t|t-1}C' + R)^{-1}(CA (x_{t-1}-\tilde{m}_{t-1}) + r_t).\nonumber
\end{eqnarray}
Thus, the pair $(\tilde{m}_t,\Sigma_{t|t-1})$ is a Markov source, where the evolution of $\Sigma_{t|t-1}$ is deterministic. Even though the cost to be minimized is not bounded, since $\tilde{m}_t$ itself is a fully observed process, the proof of Theorem \ref{thm1} can be modified to develop the structural result that any causal encoder can be replaced with one which uses $(\tilde{m}_t, \Sigma_{t|t-1})$ and the past quantization outputs (this result can be proven also using Theorem 1 of Witsenhausen \cite{Witsenhausen}, since this source is fully observed by the encoder). Likewise, the proof of Theorem \ref{thm3} shows that, for the fully observed Markov source $(\tilde{m}_t,\Sigma_{t|t-1})$, any causal coder can be replaced with one which only uses the conditional probability on $\tilde{m}_t$ and the realization $(\tilde{m}_t,\Sigma_{t|t-1},t)$ at time t. \hfill $\diamond$


Thus, the optimality of Kalman Filtering allows the encoder to only use the conditional estimate and the error covariance matrix without any loss of optimality (See Figure \ref{KalmanOptimal}), and the optimal quantization problem also has an explicit formulation. The above result is related to findings in \cite{Tsybakov} (also see \cite{Ayanoglu} and \cite{Fine}), and partially improves them in the direction of Markov sources.

We note that, the above result also applies to the settings when a controller acts on the system, that is, there exists for $u_t \in \mathbb{R}^m$ and a matrix $B$ such that $x_{t+1} = Ax_t + Bu_t + w_t$. In this case, the well-known principle of separation and control in control theory allows the above results to be applicable. In particular, the conditional estimation error is not affected by the control actions.

\section{Conclusion}

For the optimal causal coding of a partially observed Markov source, the structure of the optimal causal coders is obtained and is shown to admit a separation structure. We observed in particular that {\em separation of estimation (conditional probability computation) and quantization (of this probability)} applies, under such a setup. We also observed that the real-time decentralized coding of a partially observed i.i.d. source admits a separation. Such a separation result does not, in general, extend to decentralized coding of partially observed Markov sources.

The results and the general program presented in this paper apply also to coding over discrete memoryless noisy channels with noiseless feedback. We note that Walrand and Varaiya \cite{WalrandVaraiya} considered the noisy channel setting in their analysis in the presence of noiseless feedback.

The separation result will likely find many applications in sensor networks and networked control problems where sensors have imperfect observation of a plant to be controlled. One direction is to find explicit results on the optimal policies using computational tools. One promising approach is the expert-based systems, which are very effective once one imposes a structure on the designs, see \cite{GyorgyLinder2} for details.

One further problem is on the existence and design of optimal quantizers. Existence of optimal quantizers, even in the context of vector quantization for $\mathbb{R}^n$-valued random variables, requires stringent conditions. Such proofs typically have the form of Weierstrass theorem: A lower semi-continuous function over a compact set admits a minimum. Existence results for optimal quantizers for a one-stage cost problem has been investigated by Abaya and Wise \cite{AbayaWise} and Pollard \cite{Pollard} for continuous cost functions which are non-decreasing in the source-reconstruction distance. \cite{YukselOptimizationofChannels} obtained existence results for more general cost functions under the restriction that the code bins/cells are convex and the source admits a density function. For dynamic quantizers, \cite{YukLinCDC2012} established the existence of optimal quantization policies under the assumption that the quantizers admit the structure suggested by Theorem \ref{thm3} for fully observed Markov sources and the codecells are convex for a class of Markov sources. Also for dynamic vector quantizers, \cite{Borkar} investigated the existence results when there is a bound on the quantizer bins.

Theorem \ref{thm3} motivates the problem of optimal quantization of probability measures. This remains as an interesting problem to be investigated in a real-time coding context, with important practical consequences in control and economics applications. With a separation result paving the way for an MDP formulation, one could proceed with the analysis of \cite{Borkar} with the evaluation of optimal quantization policies and existence results for infinite horizon problems. Toward this direction, Graf and Luschgy, in \cite{Graf} and \cite{Graf2}, have studied the optimal quantization of probability measures.

One related question to be pursued further is the following: When is there an incentive for signaling in coding problems? When the observations are correlated for sources with memory or when the real-time coding of possibly independent sources over a general MAC type channel is considered, there may be an incentive for signaling. Further results on this will provide some light on some outstanding problems such as the capacity of MAC channels with feedback \cite{CoverLeung}.

\appendix

\subsection{Proof of Theorem \ref{thm1}}

We transform the problem into a real-time coding problem involving a fully observed Markov source.
At time $t=T-1$, the per-stage cost function can be written as follows, where $\gamma^0_t$ denotes a fixed receiver policy:
\begin{eqnarray}\label{costCompactForm}
&&E[c(x_{t},\gamma^0_{t}(q_{[0,t]})) | q_{[0,t-1]}]  \nonumber \\
&& = \sum_{{\cal M}_t} P(q_t=k | q_{[0,t-1]}) \bigg(\int_{\mathbb{X}} P(dx_{t}|q_{[0,t-1]},k) \nonumber \\
&& \quad \quad\quad \quad \quad \quad \times c(x_{t},\gamma^0_t(q_{[0,t-1]},k))\bigg) \nonumber \\
&& = \int_{\mathbb{X}} \sum_{{\cal M}_t} P(dx_{t}, q_t=k |q_{[0,t-1]}) c(x_{t},\gamma^0_t(q_{[0,t-1]},k))\nonumber \\
&& = \int_{{\cal P}(\mathbb{X})} \int_{\mathbb{X}} \sum_{{\cal M}_t} P(dx_{t}, q_t=k , d\pi_t |q_{[0,t-1]}) \nonumber \\
&& \quad \quad \quad \quad \times c(x_{t},\gamma^0_t(q_{[0,t-1]},k))\nonumber \\
&& = \int_{{\cal P}(\mathbb{X})} \int_{\mathbb{X}} \sum_{{\cal M}_t}  P(d\pi_t|q_{[0,t-1]}) \nonumber \\
&& \quad \quad \times P(dx_t| \pi_t) P(q_t=k | \pi_t , q_{[0,t-1]})  c(x_{t},\gamma^0_t(q_{[0,t-1]},k)) \nonumber \\
&& = \int_{{\cal P}(\mathbb{X})} \sum_{{\cal M}_t} P(d\pi_t|q_{[0,t-1]}) P(q_t=k | \pi_t , q_{[0,t-1]}) \nonumber \\
&& \quad \quad \quad \quad \times \int_{\mathbb{X}} P(dx_t| \pi_t)   c(x_{t},\gamma^0_t(q_{[0,t-1]},k)) \nonumber \\
&& = E[F(\pi_t,q_{[0,t-1]},q_t) | q_{[0,t-1]}],
\end{eqnarray}
where, $\pi_t(\cdot)=P(x_t \in \cdot|y_{[0,t]})$ and
\[F(\pi_t,q_{[0,t-1]},q_t) = \int_{\mathbb{X}} \pi_t(dx) c(x,\gamma^0_t(q_{[0,t-1]},q_t)).\]

In the above derivation, the fourth equality follows from the property that \[x_t \leftrightarrow P(dx_t|y_{[0,t]}) \leftrightarrow q_{[0,t]} \]

We note that $F(\cdot,\gamma^0_t(q_{[0,t-1]},q_t))$ is measurable on ${\cal P(\mathbb{X})}$ under weak convergence topology by Theorem \ref{MonotoneClassRamon} and the fact that the cost is measurable and bounded.

It should be noted that for every composite quantization policy, one may define $q_t$ as a random variable on the probability space such that the joint distribution of $(q_t,\pi_t,q_{[0,t-1]})$ matches the characterization that $q_t=Q^{comp}_t(y_{[0,t]},q_{[0,t-1]})$, since
\begin{eqnarray}
&&P(q_t|\pi_t,q_{[0,t-1]}) \nonumber \\
&&= \int_{\mathbb{Y}^{t+1}} P(q_t|y_{[0,t]},q_{[0,t-1]})) P(dy_{[0,t]} |\pi_t,q_{[0,t-1]}). \nonumber
\end{eqnarray}
The final stage cost is thus written as \[E[F(\pi_t,q_{[0,t-1]},q_t) | q_{[0,t-1]}],\]
which is equivalent to, by the smoothing property of conditional expectation, the following:
$$E\bigg[E [F(\pi_t,q_{[0,t-1]},q_t) | \pi_t,q_{[0,t-1]}]  \bigg| q_{[0,t-1]}\bigg]$$

Now, we will apply Witsenhausen's two stage lemma \cite{Witsenhausen}, to show that we can obtain a lower bound for the double expectation by picking $q_t$ as a result of a measurable function of $\pi_t,q_{[0,t-1]}$. Thus, we will find a composite quantization policy which only uses $(\pi_t,q_{[0,t-1]})$ which performs as well as one which uses the entire memory available at the encoder. To make this precise, let us fix the decision function $\gamma^0_t$ at the receiver corresponding to a given composite quantization policy at the encoder $Q_t^{comp}$, let $t=T-1$, and define for every $k \in {\cal M}_t$:
\begin{eqnarray}
&& \beta_{k} := \bigg\{\pi_t,q_{[0,t-1]} : F(\pi_t,q_{[0,t-1]},k) \nonumber \\
&& \quad \quad \leq F(\pi_t,q_{[0,t-1]},q'), \forall q'\neq k, q' \in {\cal M}_t \bigg\}. \nonumber
\end{eqnarray}
These sets are Borel, by the measurability of $F$ on ${\cal P}(\mathbb{X})$. Such a construction covers the domain set consisting of $(\pi_t,q_{[0,t-1]})$ but with overlaps. It covers the elements in ${\cal P}(\mathbb{X}) \times \prod_{t=0}^{T-2} {\cal M}_{t}$, since for every element in this product set, there is a minimizing $k \in {\cal M}_t$ (note that ${\cal M}_t$ is finite). To avoid the overlaps, we adopt the following technique which was introduced in Witsenhausen \cite{Witsenhausen}. Let there be an ordering of the elements in ${\cal M}_{t}$ as $1, 2, \dots, |{\cal M}_{t}|$, and for $k \geq 1$ in this sequence define a function $Q_t^{comp,*}$ as: 
\begin{eqnarray}
&& q_t = Q_t^{comp,*}(\pi_t,q_{[0,t-1]})= k, \nonumber \\
&&\quad \quad \quad \quad \rm{if} \quad (\pi_t,q_{[0,t-1]}) \in \beta_{k} \setminus \cup_{i=1}^{k-1} \beta_{i}, \nonumber
\end{eqnarray}
with $\beta_0=\emptyset.$
Thus, for any random variable $q_t$ appropriately defined on the probability space,
\begin{eqnarray}
&&E\bigg[E [F(\pi_t,q_{[0,t-1]},q_t) | \pi_t,q_{[0,t-1]}]  \bigg| q_{[0,t-1]}\bigg] \nonumber \\
&& \geq E\bigg[E [F(\pi_t,q_{[0,t-1]},Q_t^{comp,*}(\pi_t,q_{[0,t-1]}) ) | \pi_t,q_{[0,t-1]}]  \nonumber \\
&& \quad \quad \quad \quad \quad \quad \quad \quad \quad \quad \quad \bigg| q_{[0,t-1]}\bigg] \nonumber
\end{eqnarray}
Thus, the new composite policy performs at least as well as the original composite coding policy even though it has a restricted structure.

As such, any composite policy can be replaced with one which uses only $\{\pi_t,q_{[0,t-1]}\}$ without any loss of performance, while keeping the receiver decision function $\gamma^0_t$ fixed. It should now be noted that  $\{\pi_t\}$ is a Markov process: Note that the following filtering equation applies \cite{BorkarBudhiraja}
\begin{eqnarray}
&& P(dx_t|dy_{[0,t]}) \nonumber \\
&&= {\int_{x_{t-1}} P(dy_t|x_t)P(dx_t|x_{t-1})P(dx_{t-1}|dy_{[0,t-1]}) \over \int_{x_{t-1},x_{t}} P(dy_t|x_t)P(dx_t|x_{t-1})P(dx_{t-1}|dy_{[0,t-1]})}, \nonumber
\end{eqnarray}
and $P(dy_t| \pi_{s},s \leq t-1) = \int_{x_t} P(dy_t,dx_t | \pi_{s},s \leq t-1) = P(dy_t| \pi_{t-1})$. These imply that (see \cite{BorkarBudhiraja}, see also the discussion following (\ref{whySeparationHolds})) the following defines a Markov kernel (that is, a regular conditional probability measure):
\begin{eqnarray}\label{MarkovBelief}
P(d\pi_t | \pi_s,s \leq t-1 ) = P(d\pi_t| \pi_{t-1}).
\end{eqnarray}
We have thus obtained the structure of the encoder for the final stage. We iteratively proceed to study the other time stages. 
In particular, since $\{\pi_t\}$ is Markov, we could proceed as follows (in essence using Witsenhausen's three-stage lemma \cite{Witsenhausen}): For a three-stage cost problem, the cost at time $t=2$ can be written as, for measurable functions $c_2, c_3$:
\begin{eqnarray}
&& E\bigg[ c_2(\pi_2,v_2(q_{[1,2]}),q_{[1,2]}) \nonumber \\
&& + E[c_3(\pi_3,v_3(q_{[1,2]},Q^{comp,*}_3(\pi_3,q_{[1,2]})))|q_{[1,2]},\pi_{[1,2]}]  \nonumber \\
&& \quad \quad \quad \quad \quad \quad \quad \quad \bigg| q_{[1,2]},\pi_{[1,2]} \bigg] \nonumber
\end{eqnarray}
Since \[P(d\pi_3,q_2,q_1|\pi_2,\pi_1,q_2,q_1)=P(d\pi_3,q_2,q_1|\pi_2,q_2,q_1),\]
 and since under $Q^{comp,*}_3$, $q_3$ is a function of $\pi_3$ and $q_1,q_2$, the expectation above is equal to, for some measurable $F_2(.)$, $E[F_2(\pi_2,q_2,q_1)|\pi_2,\pi_1,q_2,q_1]$. Following similar steps as earlier, it can be established now that a composite quantization policy at time $2$ uses $\pi_2$ and $q_1$ and is without any loss.

By a similar argument, an encoder at time $t$, $1 \leq t\leq T-1$ only uses $(\pi_t,q_{[0,t-1]})$. The encoder at time $t=0$ uses $\pi_0$, where $\pi_0=\nu_0$ is the prior distribution on the initial state.

Now that we have obtained the structure for a composite policy, we can express this as:
$$Q_t^{comp}(\pi_t,q_{[0,t-1]}) = Q^{q_{[0,t-1]}}(\pi_t), \quad \forall \pi_t,q_{[0,t-1]}$$
such that the quantizer action $Q^{q_{[0,t-1]}} \in \mathbb{Q}({\cal P}(\mathbb{X});{\cal M}_t)$ is generated using only $q_{[0,t-1]}$, and the quantizer outcome is generated by evaluating $Q^{q_{[0,t-1]}}(\pi_t)$ for every $\pi_t$. \hfill $\diamond$

\subsection{Proof of Theorem \ref{thm3}}

At time $t=T-1$, the per-stage cost function can be written as:
\begin{eqnarray}
&&E[c(x_{t},\gamma_{t}(q_{[0,t]})) | q_{[0,t]}] \nonumber \\
&&=E[\int_{\mathbb{X}} P(dx_{t}|q_{[0,t-1]},q_t) c(x_{t},\gamma_{t}(q_{[0,t]}))] \nonumber
\end{eqnarray}
Thus, at time $t=T-1$, an optimal receiver (which is deterministic without any loss of optimality, see \cite{Blackwell2}) will use $P(dx_t|q_{[0,t]})$ as a sufficient statistic for an optimal decision (or any receiver can be replaced with one which uses this sufficient statistic without any loss). Let us fix such a receiver policy which only uses the posterior $P(dx_t|q_{[0,t]})$ as its sufficient statistic. Note that
\begin{eqnarray}
P(dx_t|q_{[0,t]}) &=& \int_{\pi_t} P(dx_t|\pi_t) P(d\pi_t|q_{[0,t]}) \nonumber
\end{eqnarray}
and further
\begin{eqnarray}
&& P(d\pi_t|q_{[0,t]}) = { P(q_t,d\pi_t|q_{[0,t-1]}) \over \int_{\pi_t} P(q_t,d\pi_t|q_{[0,t-1]})}  \nonumber \\
&& \quad \quad = { P(q_t | \pi_t,q_{[0,t-1]}) P(d\pi_t|q_{[0,t-1]})  \over \int_{\pi_t} P(q_t | \pi_t, q_{[0,t-1]}) P(d\pi_t|q_{[0,t-1]}) } \label{whySeparationHolds}
\end{eqnarray}
The term $P(q_t | \pi_t , q_{[0,t-1]})$ is determined by the quantizer action $Q_t$ (this follows from Theorem \ref{thm1}).
Furthermore, given $Q_t$, the relation (\ref{whySeparationHolds}) is measurable on ${\cal P}({\cal P}(\mathbb{X}))$ (that is, in $\Xi_t(\cdot) = P(\pi_t \in \cdot|q_{[0,t-1]})$) under weak convergence.

To prove this technical argument, consider the numerator in (\ref{whySeparationHolds}) and note that the function $\kappa_B: {\cal P}({\cal P}(\mathbb{X})) \to \mathbb{R}$ defined as $\kappa_B(\Xi)=\Xi(B)$ is measurable under weak convergence topology as a consequence of Theorem \ref{MonotoneClassRamon}, for each $B \in {\cal B}({\cal P}(\mathbb{X}))$. By Theorem 2.1 in Dubins and Freedman \cite{DubinsFreedman}, this implies that the relation in (\ref{whySeparationHolds}) is measurable on ${\cal P}({\cal P}(\mathbb{X}))$ (since the topology considered in \cite{DubinsFreedman} is not stronger than the weak convergence topology, the result in \cite{DubinsFreedman} holds in this case as well).

Let us denote the quantizer applied, given the past realizations of quantizer outputs as $Q_t^{q_{[0,t-1]}}$. Note that $q_t$ is deterministically determined by $(\pi_t,Q_t^{q_{[0,t-1]}})$ and the optimal receiver function can be expressed as $\gamma^0_t(\Xi_t,q_t)$ (as a measurable function), given $Q_t^{q_{[0,t-1]}}$. The cost at time $t=T-1$ can be expressed, given the quantizer $Q_t^{q_{[0,t-1]}}$, for some Borel function $G$, as $G(\Xi_t,Q_t^{q_{[0,t-1]}})$, where
 \begin{eqnarray*}
&& G(\Xi_t,Q_t^{q_{[0,t-1]}}) \\
&& = \int_{{\cal P}(\mathbb{X})} \Xi_t(d\pi_t) \sum_{{\cal M}_t} 1_{\{q_t=Q_t^{q_{[0,t-1]}}(\pi_t)\}} \eta^{Q^{q_{[0,t-1]}}}(\Xi_t,q_t)),
 \end{eqnarray*}
with
\begin{eqnarray}
\eta^{Q^{q_{[0,t-1]}}}(\Xi_t,q_t)) = \int \pi_t(dx_t) c(x_t,\gamma^0_t(\Xi_t,q_t))
\end{eqnarray}

Now, one can construct an equivalence class among the past $q_{[0,t-1]}$ sequences which induce the same $\Xi_t$, and can replace the quantizers $Q_t^{q_{[0,t-1]}}$ for each class with one which induces a lower cost among the finitely many elements in each such class, for the final time stage. Thus, an optimal quantization output may be generated using $\Xi_t(\cdot)= P(\pi_{t} \in \cdot|q_{[0,t-1]})$ and $\pi_t$. Since there are only finitely many past sequences and finitely many $\Xi_t$, this leads to a Borel measurable selection of $\pi_t$ for every $\Xi_t$, leading to a quantizer and a measurable selection in $\Xi_t, \pi_t$.


Since such a selection for $Q_t$ only uses $\Xi_t$, an optimal quantization output may be generated using $\Xi_t(\cdot)= P(\pi_{t} \in \cdot|q_{[0,t-1]})$ and $\pi_t$. Hence, $G(\Xi_t,Q_t^{q_{[0,t-1]}})$ can be replaced with $F_t(\Xi_t)$ for some $F_t$, without any performance loss.

The same argument applies for all time stages: At time $t=T-2$, the sufficient statistic both for the immediate cost, and the cost-to-go is $P(dx_{t-1}|q_{[0,t-1]})$, and thus for the cost impacting the time stage $t=T-1$, as a result of the optimality result for $Q_{T-1}$. To show that the separation result generalizes to all time stages, it suffices to prove that $\{(\Xi_{t},Q_t)\}$ is a controlled Markov chain, if the encoders use the structure above.

Toward this end, we establish that for $t \geq 1$, for all $B \in {\cal B}({\cal P}({\cal P}(\mathbb{X})))$, equation (\ref{eqnTempHolder}) (displayed on the next page) holds.

\begin{figure*}[!t]
\normalsize
\setcounter{mytempeqncnt}{\value{equation}}
\setcounter{equation}{16}
\begin{eqnarray}
&&P\bigg( P(d\pi_t|q_{[0,t-1]}) \in B \bigg |  P(d\pi_s|q_{[0,s-1]}), Q_{s}, s \leq t-1 \bigg) \nonumber \\
&&=P\bigg( \int_{\pi_{t-1}} P(d\pi_t,d\pi_{t-1}|q_{[0,t-1]}) \in B \bigg|  P(d\pi_s|q_{[0,s-1]}), Q_{s}, s \leq t-1 \bigg) \nonumber \\
&&=P\bigg( \bigg\{ { \int_{\pi_{t-1}} P(d\pi_t|\pi_{t-1}) P(q_{t-1}|\pi_{t-1}, q_{[0,t-2]})  P(d\pi_{t-1}|q_{[0,t-2]}) \over  \int_{\pi_{t-1},\pi_t}  P(d\pi_t|\pi_{t-1}) P(q_{t-1}|\pi_{t-1}, q_{[0,t-2]})  P(d\pi_{t-1}|q_{[0,t-2]}) } \bigg\} \in B \bigg|  P(d\pi_s|q_{[0,s-1]}), Q_{s}, s \leq t-1 \bigg) \nonumber \\
&&=P\bigg( \bigg\{ { \int_{\pi_{t-1}} P(d\pi_t|\pi_{t-1}) P(q_{t-1}|\pi_{t-1}, q_{[0,t-2]})  P(d\pi_{t-1}|q_{[0,t-2]}) \over  \int_{\pi_{t-1},\pi_t} P(d\pi_t|\pi_{t-1}) P(q_{t-1}|\pi_{t-1}, q_{[0,t-2]})  P(d\pi_{t-1}|q_{[0,t-2]}) } \bigg\} \in B \bigg|  P(d\pi_{t-1}|q_{[0,{t-2}]}), Q_{t-1}\bigg) \label{SeparationPart22} \\
&&=P\bigg( \int_{\pi_{t-1}} P(d\pi_t,d\pi_{t-1}|q_{[0,t-1]}) \in B \bigg|  P(d\pi_{t-1}|q_{[0,t-2]}), Q_{t-1} \bigg) \nonumber \\
&&=P\bigg( P(d\pi_t|q_{[0,t-1]}) \in B \bigg|  P(d\pi_{t-1}|q_{[0,t-2]}), Q_{t-1} \bigg)\label{eqnTempHolder}
\end{eqnarray}
\setcounter{equation}{\value{mytempeqncnt}}
\hrulefill
\vspace*{4pt}
\end{figure*}
\addtocounter{equation}{2}

Here, (\ref{SeparationPart22}) uses the fact that $P(q_{t-1}|\pi_{t-1}, q_{[0,t-2]})$ is identified by $\{P(d\pi_{t-1}|q_{[0,{t-2}]}), Q_{t-1}\}$, which in turn is uniquely identified by $q_{[0,t-2]}$ and $Q_{t-1}$. Furthermore, the relation in (\ref{eqnTempHolder}) defines a regular conditional probability measure since for all $B \in {\cal B}({\cal P}(\mathbb{X}))$,
\begin{eqnarray}
&&\Xi_t(B) = P(\pi_t \in B|q_{[0,t-1]}) \nonumber \\
&& = \int_{\pi_{t-1}} P(\pi_t \in B,d\pi_{t-1}|q_{[0,t-1]})\nonumber \\
&& \quad = \int_{\pi_{t-1}} P(\pi_t \in B|\pi_{t-1}) P(d\pi_{t-1}|q_{[0,t-1]}) \nonumber
\end{eqnarray}
is measurable in $\Xi_{t-1}$, given $Q_{t-1}$ (as a consequence of the measurability of (\ref{whySeparationHolds}) in $\Xi_{t}$). Hence, by the result of Dubins and Freedman mentioned above (Theorem 2.1 in \cite{DubinsFreedman}), we conclude that for any measurable function $F_t$ of $\Xi_t$
\begin{eqnarray}
E[F_t(\Xi_t)| \Xi_{[0,t-1]}, Q_{[0,t-1]} ] = E[F_t(\Xi_t), Q_{t})| \Xi_{t-1}, Q_{t-1}], \nonumber
\end{eqnarray}
for every given $Q_{t-1}$. Now, once again an equivalence relationship between the finitely many past quantizer outputs, based on the equivalence of the conditional measures $\Xi_{t-1}$ they induce, can be constructed. With the controlled Markov structure, we can follow the same argument for earlier time stages. Therefore, it suffices that the encoder uses only $(\Xi_t,t)$ as its sufficient statistic for all time stages, to generate the optimal quantizer. An optimal quantizer uses $\pi_t$ to generate the optimal quantization outputs. \hfill $\diamond$

\subsection{Proof of Theorem \ref{mainDecentralized}}

The proof is in three steps, (i), (ii) and (iii) below. \\
\noindent {\bf Step (i)} In decentralized dynamic decision problems where the decision makers have the same objective (that is, in team problems), more information provided to the decision makers does not lead to any degradation in performance, since the decision makers can always choose to ignore the additional information. In view of this, let us relax the information structure in such a way that the decision makers now have access to all the previous observations, that is the information available at the encoders $1$ and $2$ are:
\[I^i_t=\{y^i_t,{\bf y}_{[0,t-1]}, {\bf q}_{[0,t-1]}\} \quad t \geq 1, \quad i=1,2.\]
\[I^i_0=\{y^i_0\}, \quad i=1,2.\]
The information pattern among the encoders is now the {\em one-step delayed observation sharing pattern}. We will show that the past information can be eliminated altogether, to prove the desired result. \\

\noindent \textbf{Step  (ii)} The second step uses the following technical Lemma.
\begin{lem}\label{lem2}
Under the relaxed information structure in step $(i)$ above, any decentralized quantization policy at time $t$, $1 \leq t\leq T-1$, can be replaced, without any loss in performance, with one which only uses $(\pi_t, {\bf y}_t,{\bf q}_{[0,t-1]})$, satisfying the following form:
\begin{eqnarray}
&& P({\bf q}_t|{\bf y}_{[0,t]},{\bf q}_{[0,t-1]}) \nonumber \\
&& =  P(q^1_t | y^1_t,{\bf q}_{[0,t-1]}) P(q^2_t | y^2_t,{\bf q}_{[0,t-1]}) \nonumber \\
&& = 1_{\{q^1_t=\bar{Q}^1(y^1_t,{\bf q}_{[0,t-1]})\}}1_{\{q^2_t=\bar{Q}^{2}(y^2_t,{\bf q}_{[0,t-1]})\}},
\end{eqnarray}
for measurable functions $\bar{Q}^1$ and $\bar{Q}^2$. \hfill $\diamond$ \\
\end{lem}
\textbf{Proof.} Let us fix a composite quantization policy ${\bf \Pi}^{comp}$. At time $t=T-1$, the per-stage cost function can be written as:
\begin{eqnarray}\label{costCompactForm2}
&& E [\int_{\mathbb{X}} P(dx_{t}|{\bf q}_{[0,t]}) c(x_{t},v_t)| {\bf q}_{[0,t-1]} ]
\end{eqnarray}
For this problem, $P(dx_{t}|{\bf q}_{[0,t]})$ is a sufficient statistic for an optimal receiver. Hence, at time $t=T-1$, an optimal receiver will use $P(dx_t|{\bf q}_{[0,t]})$ as a sufficient statistic for an optimal decision as the cost function conditioned on ${\bf q}_{[0,t]}$ is written as: $\int P(dx_t|{\bf q}_{[0,t]}) c(x_t,v_t)$, where $v_t$ is the decision of the receiver. Now, let us fix this decision policy at time $t$. We now note that (\ref{longequationDecentralized}), displayed on the next page, follows.

\begin{figure*}[!t]
\normalsize
\setcounter{mytempeqncnt}{\value{equation}}
\setcounter{equation}{20}
\begin{eqnarray}
P(dx_t|{\bf q}_{[0,t]}) &=& \sum_{\mathbb{Y}^{t+1}} P(dx_t,{\bf y}_{[0,t]}|{\bf q}_{[0,t]}) = {\sum_{\mathbb{Y}^{t+1}} P(dx_t,{\bf q}_t,{\bf y}_{[0,t]}|{\bf q}_{[0,t-1]}) \over P({\bf q}_t|{\bf q}_{[0,t-1]})} \nonumber \\
&=&{ \sum_{\mathbb{Y}^{t+1}}P({\bf q}_t|{\bf y}_{[0,t]},{\bf q}_{[0,t-1]}) P({\bf y}_t|x_t) P(dx_t|{\bf y}_{[0,t-1]}) P({\bf y}_{[0,t-1]}|{\bf q}_{[0,t-1]}) \over \int_{\mathbb{X}, \mathbb{Y}^{t+1}} P({\bf q}_t|{\bf y}_{[0,t]},{\bf q}_{[0,t-1]}) P({\bf y}_t|x_t) P(dx_t|{\bf y}_{[0,t-1]}) P({\bf y}_{[0,t-1]}|{\bf q}_{[0,t-1]}) } \nonumber \\
&=&{ \sum_{\mathbb{Y}^{t+1}}P({\bf q}_t|{\bf y}_{[0,t]},{\bf q}_{[0,t-1]}) P({\bf y}_t|x_t) \pi(dx_t) P({\bf y}_{[0,t-1]}|{\bf q}_{[0,t-1]}) \over \int_{\mathbb{X}, \mathbb{Y}^{t+1}} P({\bf q}_t|{\bf y}_{[0,t]},{\bf q}_{[0,t-1]}) P({\bf y}_t|x_t) \pi(dx_t) P({\bf y}_{[0,t-1]}|{\bf q}_{[0,t-1]}) }
\label{longequationDecentralized}
\end{eqnarray}
\setcounter{equation}{\value{mytempeqncnt}}
\hrulefill
\vspace*{4pt}
\end{figure*}
\addtocounter{equation}{1}

In (\ref{longequationDecentralized}), we use the relation $P(dx_t|{\bf y}_{[0,t-1]})=P(dx_t)=:\pi(dx_t)$, where $\pi(\cdot)$ denotes the marginal probability on $x_t$ (recall that the source is memoryless). The term $P({\bf q}_t|{\bf y}_{[0,t]},{\bf q}_{[0,t-1]})$ in (\ref{longequationDecentralized}) is determined by the composite quantization policies:
\begin{eqnarray}
&&P({\bf q}_t|{\bf y}_{[0,t]},{\bf q}_{[0,t-1]}) \nonumber \\
&&=  P(q^1_t | y^1_t,{\bf y}_{[0,t-1]},{\bf q}_{[0,t-1]})P(q^2_t | y^2_t,{\bf y}_{[0,t-1]},{\bf q}_{[0,t-1]}) \nonumber \\
&&= 1_{\{q^1_t=Q_t^{comp,1}(y^1_t,{\bf y}_{[0,t-1]},{\bf q}_{[0,t-1]})\}} \nonumber \\
&& \quad \quad \times 1_{\{q^2_t=Q_t^{comp,2}(y^2_t,{\bf y}_{[0,t-1]},{\bf q}_{[0,t-1]})\}} \nonumber
\end{eqnarray}
The above is valid since each encoder knows the past observations of both encoders.
As such, $P(dx_t|{\bf q}_{[0,t]})$ can be written as, for some function $\Upsilon$: $\Upsilon(\pi,{\bf q}_{[0,t-1]},{\bf Q}_t^{comp}(.))$. Note that ${\bf q}_{[0,t-1]}$ appears due to the term $P({\bf y}_{[0,t-1]}|{\bf q}_{[0,t-1]})$. Now, consider the following space of joint (team) mappings at time $t$, denoted by ${\cal G}_t$:
\begin{eqnarray}
{\cal G}_t=\{{\bf \Psi}_t: {\bf \Psi}_t =\{\Psi^1_t,\Psi^2_t\}, \Psi^i_t: \mathbb{Y}^i \to {\cal M}^i_t, \quad i=1,2\}. \label{spaceOfTeamActions}
\end{eqnarray}
For every composite quantization policy there exists a distribution $P'$ on random variables $({\bf q}_t, \pi, {\bf q}_{[0,t-1]})$ such that
\begin{eqnarray}\label{tulsa}
&& P'({\bf q}_t | \pi, {\bf q}_{[0,t-1]}) = \sum_{(\mathbb{Y}^1 \times \mathbb{Y}^2)^{t+1}} P({\bf q}_t ,{\bf y}_{[0,t]} | \pi, {\bf q}_{[0,t-1]})\nonumber \\
&& = \sum_{(\mathbb{Y}^1 \times \mathbb{Y}^2)^{t+1}} \bigg( P(q^1_t |{\bf y}_{[0,t-1]}, y^1_t, {\bf q}_{[0,t-1]},\pi) \nonumber \\
&& \quad  \quad \quad  \quad \quad \times P(q^2_t |{\bf y}_{[0,t-1]}, y^2_t, {\bf q}_{[0,t-1]},\pi) \nonumber \\
&& \quad  \quad \quad  \quad \quad   \times P(y^1_t,y^2_t) P({\bf y}_{[0,t-1]} |  \pi, {\bf q}_{[0,t-1]}) \bigg)
\end{eqnarray}
Furthermore, with every composite quantization policy and every realization of ${\bf y}_{[0,t-1]}, {\bf q}_{[0,t-1]}$, we can associate an element in the space ${\cal G}_t$, ${\bf \Psi}_{{\bf y}_{[0,t-1]},{\bf q}_{[0,t-1]}}$, such that the induced stochastic relationship in (\ref{tulsa}) can be obtained:
\begin{eqnarray}
&& P'({\bf q}_t | \pi, {\bf q}_{[0,t-1]}) = \sum_{(\mathbb{Y}^1 \times \mathbb{Y}^2)^{t+1}} P({\bf q}_t, {\bf y}_{[0,t]} | \pi, {\bf q}_{[0,t-1]})\nonumber \\
&=& \sum_{(\mathbb{Y}^1 \times \mathbb{Y}^2)^{t+1}} 1_{\{{\bf \Psi}_{{\bf y}_{[0,t-1]},{\bf q}_{[0,t-1]}}(y^1_t,y^2_t)=(q^1_t,q^2_t) \}} \nonumber \\
&& \quad  \quad  \quad  \quad \times P(y^1_t,y^2_t) P({\bf y}_{[0,t-1]} |  \pi,{\bf q}_{[0,t-1]}) \nonumber
\end{eqnarray}
We can thus, express the cost, for some measurable function $F$ in the following way:
$$E[F(\pi,{\bf q}_{[0,t-1]},{\bf \Psi}) | \pi,{\bf q}_{[0,t-1]}],$$
where
\begin{eqnarray}
&&P({\bf \Psi} | \pi,{\bf q}_{[0,t-1]}) \nonumber \\
&&= \sum_{(\mathbb{Y}^1 \times \mathbb{Y}^2)^{t}} 1_{ \{ {\bf \Psi} = {\bf \Psi}_{{\bf y}_{[0,t-1]},{\bf q}_{[0,t-1]}} \} } P({\bf y}_{[0,t-1]} |  \pi,{\bf q}_{[0,t-1]}) \nonumber
\end{eqnarray}
Now let $t=T-1$ and define for every realization ${\bf \Psi}_t=(\Psi^1_t,\Psi^2_t) \in {\cal G}_t$ (with the decision policy considered earlier fixed):
\begin{eqnarray}
&&\beta_{{\bf \Psi}_t} := \bigg\{\pi,{\bf q}_{[0,t-1]} : \nonumber \\
&&  \quad  \quad  \quad F(\pi,{\bf q}_{[0,t-1]},{\bf \Psi}_t) \leq F(\pi,{\bf q}_{[0,t-1]},{\bf \Psi}'_t) \nonumber \\
&& \quad  \quad  \quad  \quad  \quad  \quad  \quad  \quad  \quad  \quad  \quad \forall ((\Psi^1_t)',(\Psi^2_t)') \in {\cal G}_t \bigg\}. \nonumber
\end{eqnarray}
As we had observed in the proof of Theorem \ref{thm1}, such a construction covers the domain set consisting of $(\pi,{\bf q}_{[0,t-1]})$ but possibly with overlaps. Note that for every $(\pi,{\bf q}_{[0,t-1]})$, there exists a minimizing function in ${\cal G}_t$, since ${\cal G}_t$ is a finite set. In this sequence, let there be an ordering of the finitely many elements in ${\cal G}_t$  as $\{{\bf \Psi}_t(1),{\bf \Psi}_t(2),\dots, {\bf \Psi}_t(k),\dots\}$, and define a function ${\bf T}^*_t$ as:
\begin{eqnarray}
&&{\bf \Psi}_t(k)= {\bf T}^*_t(\pi,{\bf q}_{[0,t-1]}), \nonumber \\
&& \quad \rm{if} \quad \bigg(\pi,{\bf q}_{[0,t-1]}\bigg) \in \beta_{{\bf \Psi}_t(k)} - \cup_{i=0}^{k-1} \beta_{{\bf \Psi}_t(i)},\nonumber
\end{eqnarray} with $\beta_{{\bf \Psi}_t(0)}=\emptyset$.
Thus, we have constructed a policy which performs at least as well as the original composite quantization policy. It has a restricted structure in that it only uses $(\pi,{\bf q}_{[0,t-1]})$ to generate the team action and the local information $y^1_t,y^2_t$ to generate the quantizer outputs.

Now that we have obtained the structure of the optimal encoders for the last stage, we can sequentially proceed to study the other time stages. 
Note that given a fixed $\pi$, $\{(\pi,{\bf y}_t)\}$ is i.i.d. and hence Markov. Now, define $\pi'_t= (\pi,{\bf y}_t)$. For a three-stage cost problem, the cost at time $t=2$ can be written as, for measurable functions $c_2, c_3$:
\begin{eqnarray}
&&c_2(\pi'_2,v_2({\bf q}_{[1,2]})) \nonumber \\
&&  \quad  + E[c_3(\pi'_3,v_3({\bf q}_{[1,2]},Q_3(\pi'_3,{\bf q}_{[1,2]})))|\pi'_{[1,2]},{\bf q}_{[1,2]}] \nonumber
\end{eqnarray}
Since $P(d\pi'_3,{\bf q}_{[1,2]}|\pi'_2,\pi'_1,{\bf q}_{[1,2]})=P(d\pi'_3,{\bf q}_{[1,2]}|\pi'_2,{\bf q}_{[1,2]}),$
the expression above is equal for some $F_2(\pi'_2,{\bf q}_2,{\bf q}_1)$ for some measurable $F_2$. By a similar argument, an optimal composite quantizer at time $t$, $1 \leq t\leq T-1$ only uses $(\pi, {\bf y}_t,{\bf q}_{[0,t-1]})$. An optimal (team) policy generates the quantizers $Q^1_t,Q^2_t$ using ${\bf q}_{[0,t-1]}, \pi$, and the quantizers use $\{y^i_t\}$ to generate the quantizer outputs at time $t$ for $i=1,2$. \hfill $\diamond$ \\

\noindent {\bf Step  (iii)} The final step will complete the proof. At time $t=T-1$, an optimal receiver will use $P(dx_t|{\bf q}_{[0,t]})$ as a sufficient statistic for the optimal decision. We now observe that
\begin{eqnarray}
P(dx_t|{\bf q}_{[0,t]}) &=& \sum_{\mathbb{Y}^{t+1}} P(dx_t|{\bf y}_{[0,t]}) P({\bf y}_{[0,t]}|{\bf q}_{[0,t]}) \label{BeliefOnBelief} \\
&=& \sum_{\mathbb{Y}^{t+1}} P(dx_t|{\bf y}_{t}) P({\bf y}_{[0,t]}|{\bf q}_{[0,t]}) \nonumber \\
&=& \sum_{\mathbb{Y}} P(dx_t|{\bf y}_{t}) \sum_{\mathbb{Y}^{t}} P({\bf y}_{[0,t]}|{\bf q}_{[0,t]}) \nonumber\\
&=& \sum_{\mathbb{Y}} P(dx_t|{\bf y}_{t}) P({\bf y}_{t}|{\bf q}_{[0,t]}) \nonumber
\end{eqnarray}
Thus, $P(dx_t|{\bf q}_{[0,t]})$, is a function of $P({\bf y}_t|{\bf q}_{[0,t]})$. Now, let us note that
\begin{eqnarray}
&& P({\bf y}_t|{\bf q}_{[0,t]}) = { P({\bf q}_t,{\bf y}_t|{\bf q}_{[0,t-1]}) \over \sum_{{\bf y}_t} P({\bf q}_t,{\bf y}_t|{\bf q}_{[0,t-1]})} \nonumber\\
&& = { P({\bf q}_t | {\bf y}_t, {\bf q}_{[0,t-1]}) P({\bf y}_t|{\bf q}_{[0,t-1]})  \over \sum_{{\bf y}_t} P({\bf q}_t | {\bf y}_t, {\bf q}_{[0,t-1]}) P({\bf y}_t|{\bf q}_{[0,t-1]}) } \nonumber\\
&& = { P({\bf q}_t | {\bf y}_t, {\bf q}_{[0,t-1]}) P({\bf y}_t)  \over \sum_{{\bf y}_t} P({\bf q}_t | {\bf y}_t, {\bf q}_{[0,t-1]}) P({\bf y}_t)  },
\end{eqnarray}
where the term $P({\bf q}_t | {\bf y}_t , {\bf q}_{[0,t-1]})$ is determined by the quantizer team action ${\bf Q}_t^{comp}$. As such, the cost at time $t=T-1$ can be expressed as a measurable function $G(P({\bf y}_{t}),{\bf Q}_t)$. Thus, it follows that, an optimal quantizer policy at the last stage, $t=T-1$ may only use $P({\bf y}_{t})$ to generate the quantizers, where the quantizers use the local information $y^i_t$ to generate the quantization output. At time $t=T-2$, the sufficient statistic for the cost function is $P(dx_{t-1}|{\bf q}_{[0,t-1]})$ both for the immediate cost, and the cost-to-go, that is the cost impacting the time stage $t=T-1$, as a result of the optimality result for $Q_{T-1}$ and the memoryless nature of the source dynamics. The same argument applies for all time stages. 

Hence, any policy without loss can be replaced with one in $\Pi^{NSM}$ defined in (\ref{piNSM}). Since there are finitely many policies in this class, an optimal policy exists. \hfill $\diamond$

\textbf{Acknowledgements}
Discussions with Professors Ramon van Handel, Tam\'as Linder and Naci Saldi on the contents of the paper are gratefully acknowledged. The detailed reviews of two anonymous reviewers and the suggestions of the associate editor have led to significant improvement in the presentation.

\begin{IEEEbiographynophoto}
{Serdar Y\"uksel} received his BSc degree in Electrical and Electronics
Engineering from Bilkent University in 2001; MS and PhD degrees in
Electrical and Computer Engineering from the University of Illinois at
Urbana-Champaign in 2003 and 2006, respectively. He was a post-doctoral
researcher at Yale University for a year before joining Queen's University
as an assistant professor of Mathematics and Engineering at the Department
of Mathematics and Statistics. His research interests are on stochastic
and decentralized control, information theory and applied probability.
\end{IEEEbiographynophoto}
\end{document}